\documentclass[showpacs,showkeys,floats,aps,nofootinbib,preprint]{revtex4}
\usepackage{bm,amsfonts, mathtools}
\usepackage{graphicx,color, wasysym}
\usepackage{textcomp}
\usepackage{amsmath,amssymb,latexsym,epsfig}

\def\nn{\nonumber}

\begin{document}
\thispagestyle{empty}

\title{Dirac factorization and fractional calculus}

\author{D. Babusci}
\email{danilo.babusci@lnf.infn.it}
\affiliation{INFN - Laboratori Nazionali di Frascati, Via E. Fermi, 40, 00044 
Frascati (Roma), Italy} 
\author{G. Dattoli}
\email{giuseppe.dattoli@enea.it}
\author{M. Quattromini}
\email{marcello.quattromini@enea.it}
\affiliation{ENEA -  Centro Ricerche Frascati, Via E. Fermi, 45, 00044 
Frascati (Roma), Italy}
\author{P. E. Ricci}
\email{paoloemilioricci@gmail.com}
\affiliation{International Telematic University UniNettuno, 
Corso V. Emanuele II, 39 00186 Roma, Italy}

\begin{abstract}
We show that the Dirac factorization method can be successfully employed to treat problems involving operators raised to a fractional power. 
The technique we adopt is based on an extension of the Pauli matrices and the properties of the roots of unity. We also comment about the 
possibility of using the method to linearize evolution equations containing the $n$-th root of differential operators and make a comparison with 
other techniques involving suitable transforms.
\end{abstract}

\maketitle

The formulation of the relativistic Dirac equation \cite{Dirac} is based on a factorization of the square root that, as a consequence of the anti-commuting 
nature of Pauli matrices \cite{Louisell}
\begin{equation}
%\label{ }
\left\{\hat{\sigma}_j,\hat{\sigma}_k\right\} \,=\, 2\,\delta_{jk} \qquad\qquad (j, k = 1, 2, 3), 
\end{equation}
allows to write the Pythagorean theorem in the form 
\begin{equation}
\label{pytha}
\sqrt{A^2 + B^2} = A\,\hat{\sigma}_j + B\,\hat{\sigma}_k \qquad\qquad (j \neq k),
\end{equation}
where $A$, $B$ can be either numbers or commuting operators. Eq. \eqref{pytha} can be extended to the case of three terms, and one can write
\begin{equation}
%\label{}
\sqrt{A^2 + B^2 + C^2} = A\,\hat{\sigma}_j + B\,\hat{\sigma}_k + C\,\hat{\sigma}_l \qquad\qquad (j \neq k \neq l),
\end{equation}
The realization of such a factorization is not unique. The Hamilton quaternions \cite{Conway} can be efficiently used for this purpose, and this is by no 
means surprising if we note that a purely imaginary quaternion
\begin{equation}
%\label{ }
\hat{Q} = a\,\hat{i} + b\,\hat{j} + c\,\hat{k}, \qquad\qquad \hat{i}^2 = \hat{j}^2 = \hat{k}^2 = \hat{i}\,\hat{j}\,\hat{k} = - 1,
\end{equation}
can be realized in terms of Pauli matrices since one can write 
\begin{equation}
%\label{ }
\hat{i} = i\,\hat{\sigma}_1, \qquad \hat{j} = i\,\hat{\sigma}_2, \qquad \hat{k} = i\,\hat{\sigma}_3.
\end{equation}
It is therefore evident that the key tool of the game is the use of the Clifford algebras, at least if we limit ourselves to the case of square roots.

We believe that a further element of interest in the use of this factorization method, and in its extension to higher-order roots, is that it can open new 
perspectives in the theory of the fractional calculus \cite{OldSpa}, i. e.,  in the body of computations aimed at allowing the handling of operators (differential 
or not) raised to a non-integer number. Just to give an example, we note that, by choosing $(j, k, l) = (1, 2, 3)$, we can write 
\begin{equation}
%\label{ }
\sqrt{\nabla^2} = \left(
\begin{array}{cc}
 \partial_z   &  \partial_x - i\,\partial_y  \\
  \partial_x + i\,\partial_y     &  - \partial_z  
\end{array}
\right),
\end{equation}
while the operator $O_{1/2} = \sqrt{\partial_x + a}$ can be cast in the form
\begin{equation}
%\label{ }
O_{1/2} = \sqrt{\partial_x}\,\hat{\sigma}_j + \sqrt{a}\,\hat{\sigma}_k
\end{equation}
i.e., using the well-known identity \cite{OldSpa}
\begin{equation}
%\label{ }
\partial_x^{\nu}\,x^{\mu} = \frac{\Gamma (\mu + 1)}{\Gamma (\mu - \nu + 1)}\,x^{\mu - \nu} \qquad\qquad (\mu, \nu \in \mathbb{R}),
\end{equation}
we get
\begin{equation}
%\label{ }
O_{1/2} = \frac1{\sqrt{\pi\,x}}\,\hat{\sigma}_j + \sqrt{a}\,\hat{\sigma}_k.
\end{equation}
For example, in the case $j = 3, k = 2$ one has 
\begin{equation}
%\label{ }
O_{1/2} =  \frac1{\sqrt{\pi\,x}}\,\left(
\begin{array}{cc}
 1   &  - i\,\sqrt{\pi\,a\,x}\\
i\,\sqrt{\pi\,a\,x}   &  - 1
\end{array}
\right),
\end{equation}
or, by choosing  $j = 1, k = 2$,
\begin{equation}
\label{Op12}
O_{1/2} =  \frac1{\sqrt{\pi\,x}}\,\left(
\begin{array}{cc}
 0   &  1 - i\,\sqrt{\pi\,a\,x}\\
1 + i\,\sqrt{\pi\,a\,x}   &  0
\end{array}
\right).
\end{equation} 

In a previous note \cite{Babusci} we have shown an extension of the Dirac factorization method allows the extension of the identity \eqref{pytha} to cubic 
roots, namely
\begin{equation}
\label{cubic}
\sqrt[3]{A^3 + B^3} = A\,\hat{\lambda}_1 + B\,\hat{\lambda}_2
\end{equation}
where 
\begin{equation}
%\label{ }
\hat{\lambda}_1 = \left(
\begin{array}{ccc}
   0  &  \epsilon_0  &  0  \\
   0  &  0  &   \epsilon_0  \\
   \epsilon_0 & 0 & 0
\end{array}
\right), \qquad\qquad 
\hat{\lambda}_2 = \left(
\begin{array}{ccc}
   0  &  \epsilon_1  &  0  \\
   0  &  0  &   \epsilon_2  \\
   \epsilon_0 & 0 & 0 
\end{array}
\right), 
\end{equation}
with $\epsilon_k = e^{i\,2\,\pi\,k/n} (k = 0, 1, 2)$ cubic roots of unity satisfying the conditions
\begin{equation}
%\label{ }
\sum_{k = 0}^2 \epsilon_k = 0, \qquad \qquad \prod_{k = 0}^2 \epsilon_k = \epsilon_0 = 1,
\end{equation}
from which we can show the following identities, necessary for the fulfillment of Eq. \eqref{cubic}, hold
\begin{equation}
\label{lamid}
\hat{\lambda}_j^3 = \hat{1}, \qquad\qquad \hat{\lambda}_j^2\,\hat{\lambda}_k + \hat{\lambda}_j\,\hat{\lambda}_k\,\hat{\lambda}_j + 
\hat{\lambda}_k\,\hat{\lambda}_j^2 = 0 \qquad (j \neq k).
\end{equation}
As an example of application, in analogy to Eq. \eqref{Op12}, we can write
\begin{align}
%\label{ }
O_{1/3} = \sqrt[3]{\partial_x + a} &= \frac1{\Gamma (2/3)\,\sqrt[3]{x}}\,\hat{\lambda}_1 + \sqrt[3]{a}\,\hat{\lambda}_2 \nn \\
& = \left(
\begin{array}{ccc}
   0  &  \displaystyle \frac1{\Gamma (2/3)\,\sqrt[3]{x}} + \epsilon_1\,\sqrt[3]{a}  &  0  \\
   0  &  0  &  \displaystyle \frac1{\Gamma (2/3)\,\sqrt[3]{x}} + \epsilon_2\,\sqrt[3]{a}  \\
    \displaystyle \frac1{\Gamma (2/3)\,\sqrt[3]{x}} + \sqrt[3]{a} & 0 & 0 
\end{array}
\right), 
\end{align}

By adding to these matrices the diagonal one
\begin{equation}
%\label{ }
\hat{\lambda}_3 = \left(
\begin{array}{ccc}
   \epsilon_1  &  0  & 0  \\
   0  &  \epsilon_2  & 0  \\
   0 & 0 & 1  
\end{array}
\right), 
\end{equation}
we can realize the factorization
\begin{equation}
\label{3pyt}
\sqrt[3]{A^3 + B^3 + C^3} = A\,\hat{\lambda}_j + B\,\hat{\lambda}_k + C\,\hat{\lambda}_l \qquad\qquad (j \neq k \neq l = 1, 2, 3). 
\end{equation}
As an example, by choosing $(j, k, l) = (1, 2, 3)$ in this equation, one obtains 
\begin{equation}
%\label{ }
\sqrt[3]{\partial_x^3 + \partial_y^3 + \partial_z^3} = \left(
\begin{array}{ccc}
   \epsilon_1\,\partial_z  &  \partial_x + \epsilon_1\,\partial_y  &  0  \\
   0  &  \epsilon_2\,\partial_z  &  \partial_x + \epsilon_2\,\partial_y\\
   \partial_x + \partial_y & 0 & \partial_z
\end{array}
\right).
\end{equation}
The matrices $\hat{\lambda}_{1,2,3}$ represent a generalization of Pauli matrices, but they are not the only possible triples. For example, we can pick the 
following three matrices 
\begin{equation}
%\label{ }
\hat{\phi}_1 = \left(
\begin{array}{ccc}
   0  &  0 & \epsilon_1  \\
   \epsilon_2 & 0 &  0  \\
    0 &  1 & 0
\end{array}
\right), \qquad 
\hat{\phi}_2 = \left(
\begin{array}{ccc}
   0  &  0  &  1 \\
   1  &  0  &  0  \\
   0  &  1  &  0 
\end{array}
\right), \qquad 
\hat{\phi}_3 = \left(
\begin{array}{ccc}
   \epsilon_2  &  0  &  0  \\
   0  &  \epsilon_1  &  0   \\
   0  &  0  &   1
\end{array}
\right).
\end{equation}
In general, 24 triples are possible. A complete list of them can be found in Ref. \cite{Herrmann}\footnote{Going deeper into their algebraic structure it can be 
shown that they give a representation of SU(3) algebra generators.}, where an analogous problem has been treated within the context of the study of the fractional 
Pauli equation. The matrices $\hat{\phi}_j$ are not a generalization of Pauli matrices, as it is clear from the computation of their commutation relations 
\begin{align}
\label{comm}
[\hat{\phi}_1, \hat{\phi}_2]  &= (\epsilon_1 - 1)\,\left(
\begin{array}{ccc}
   0  &  1  &  0  \\
   0  &  0  & \epsilon_1  \\
   \epsilon_2  &  0  & 0 
\end{array}
\right) \nn \\
[\hat{\phi}_1, \hat{\phi}_3]  &= (\epsilon_1 - 1)\,\left(
\begin{array}{ccc}
   0  &  0  &  1   \\
   1  &  0  &  0  \\
   0  &  1  &  0 
\end{array}
\right) = (\epsilon_1 - 1)\,\hat{\phi}_2 \\
[\hat{\phi}_2, \hat{\phi}_3]  &= (\epsilon_1 - 1)\,\left(
\begin{array}{ccc}
   0  &  0  &  \epsilon_2  \\
   \epsilon_1  &  0  &  0  \\
   0  &  1  &  0 
\end{array}
\right). \nn
\end{align}
We remark that the matrices in r.h.s. of Eq. \eqref{comm} do not provide a suitable triple for the cubic factorization.

For the case of quartic root, we have found that the factorization occurs through a set of matrices $4 \times 4$ expressible through the Pauli matrices, and 
an example is given below 
\begin{equation}
%\label{ }
\hat{\chi}_1 = \left(
\begin{array}{cc}
   \hat{\sigma}_+  &  \hat{\sigma}_-  \\
   \hat{\sigma}_-  &   \hat{\sigma}_+ 
\end{array}
\right), \qquad
\hat{\chi}_2 = e^{\,i\pi/4}\,\left(
\begin{array}{cc}
    i\,\hat{\sigma}_+  &  - \hat{\sigma}_-  \\
   \hat{\sigma}_-  &  - i\,\hat{\sigma}_+ 
\end{array}
\right), \qquad
\hat{\chi}_3 = \left(
\begin{array}{cc}
   0  &  \hat{s}  \\
   - \hat{s}  & 0 
\end{array}
\right),
\end{equation}
where
\begin{equation}
\hat{\sigma}_+ = \left(
\begin{array}{cc}
    0  &  1  \\
    0  &  0 
\end{array}
\right), \qquad 
\hat{\sigma}_- = \left(
\begin{array}{cc}
    0  &  0  \\
    1  &  0 
\end{array}
\right), \qquad
\hat{s} = \left(
\begin{array}{cc}
   - i  &  0  \\
    0  &  1 
\end{array}
\right).
\end{equation}
We note that $[\hat{\chi}_1, \hat{\chi}_2] = \sqrt{2}\,\hat{\chi}_3$. We do not report all the matrices allowing the fourth order factorization. They can be derived 
by following a procedure analogous to that outlined for the case $n = 3$.

The case of the Dirac factorization for a generic $n$, namely
\begin{equation}
%\label{ }
\sqrt[n]{A^n + B^n} = A\,\hat{\tau}_1 + B\,\hat{\tau}_2,
\end{equation}
can be afforded using the same strategy as before, i. e., searching for the associated matrices by means of the properties of the roots of unity. In the case of $n$ 
odd we find\footnote{For $n$ even the matrix $\hat{\tau}_2$ should be replaced by $e^{\,i\,2\,\pi/n}\,\hat{\tau}_2$.}
\begin{equation}
%\label{ }
\hat{\tau}_1 = \left(
\begin{array}{ccccc}
    0  &  \epsilon_0  &  0  & \cdots  &  0  \\
    0  &  0 &  \epsilon_0   & \cdots  &  0  \\
    \vdots & \vdots & \vdots & \vdots & \vdots \\
    0  &  0  &  0 & \cdots  &  \epsilon_0 \\
    \epsilon_0 & 0 & 0 & \vdots & 0
\end{array}
\right), \qquad\qquad
\hat{\tau}_2 = \left(
\begin{array}{ccccc}
    0  &  \epsilon_1  &  0  & \cdots  &  0  \\
    0  &  0 &  \epsilon_2   & \cdots  &  0  \\
    \vdots & \vdots & \vdots & \vdots & \vdots \\
    0  &  0  &  0 & \cdots  &  \epsilon_{n - 1} \\
    \epsilon_0 & 0 & 0 & \vdots & 0
\end{array}
\right), 
\end{equation}
where $\epsilon_k\,\,(k = 0, \cdots, n - 1)$ are the $n$-th roots of unity, satisfying the conditions 
\begin{equation}
%\label{ }
\sum_{k = 0}^{n - 1} \epsilon_k = 0, \qquad \qquad \prod_{k = 0}^{n - 1} \epsilon_k = \epsilon_0 = 1,
\end{equation}
that determine identities of the type \eqref{lamid} on which is based the factorization also for this more general case. As in the previous case the matrices 
$\hat{\tau}_{1,2}$ are not the only ones enabling the factorization. 

The following example, concerning the solution of a fractional evolution equation, shows how other matrices can naturally enter into the game. A fairly immediate 
consequence of the factorization method is that we can ``linearize" the equation 
\begin{equation}
\label{Func}
\partial_t^n\,F (x, t) = \sqrt[n]{\partial_x^n + a^n}\,F (x, t)
\end{equation}
to obtain the following Dirac-like form 
\begin{equation}
%\label{ }
\partial_t\,\Phi (x, t) = (\partial_x\,\hat{\tau}_1 + a\,\hat{\tau}_2)\,\Phi (x, t),
\end{equation}
where $\Phi$ is a $n$-components vector. By letting $\Phi_0 (x) = \Phi (x, 0)$, the solution of this equation can be written as follows
\begin{equation}
%\label{ }
\Phi (x, t) = U (t)\,\Phi_0 (x), \qquad\qquad U (t) = e^{\,t(\hat{\tau}_1\,\partial_x + a\,\hat{\tau}_2)}.
\end{equation}
Since the evolution operator $U (t)$ is an exponential whose argument contains non-commuting matrices, its disentangling requires an ordered product of 
exponentials involving successive commutators of the matrices itself. At the lowest order in the commutation chain we find\footnote{The complete disentanglement 
can be achieved taking into account that the matrices $\hat{\tau}_j$ provide a representation of SU(N) algebra generators.}
\begin{equation}
%\label{ }
U (t) \cong e^{\,t\,\hat{\tau}_1\,\partial_x}\,e^{\,t\,a\,\hat{\tau}_2}\,e^{- (a\,t^2/2)\,[\hat{\tau}_1, \hat{\tau}_2]\,\partial_x}.
\end{equation} 

As discussed before (see Eq. \eqref{Func}), the Dirac factorization method can play important role in the solution of functional type equations. We want to 
emphasize that a great deal of activity has been pursued in this field and that other methods can be exploited . To clarify this point we consider the following 
generalized ``heat equation" 
\begin{equation}
\label{gheat}
\partial_t\,F (x, t) = - \sqrt[n]{- \partial_x^2 + k}\,F (x, t), \qquad\qquad F (x, 0) = f (x).
\end{equation}
By taking into account the identity \cite{Gorska}
\begin{equation}
\exp\left(- c\,p^{\nu}\right) = \int_0^\infty \mathrm{d}\xi\,g_{\nu} (\xi)\,\exp(- c^{1/\nu}\,p\,\xi) \qquad\qquad (p > 0, c > 0)  
\end{equation}
where the functions $g_\nu (\xi)$ are the so called one-sided Levy stable distribution, the solution of Eq. \eqref{gheat} can be written as
\begin{equation}
%\label{ }
F (x, t) = \int_0^\infty \mathrm{d}\xi\,g_{1/n} (\xi)\,\exp\left(- t^n\,k\,\xi\right)\,\left[\exp\left(t^n\,\xi\,\partial_x^2\right)\,f (x)\right], 
\end{equation}
where the part within square brackets is the kernel of the ordinary heat equation.

In this paper we have shown how the Dirac method can be exploited to deal with fractional operators. We have also indicated the relevant importance to treat 
equations involving higher-order square root of differential operators, and provided the link with a method exploiting the Levy distributions. We believe that the topics 
treated in this note are interesting enough to stimulate new suggestions in fractional calculus, worth to be pursued. 

\end{document}